\documentclass[prl,twocolumn,showpacs,preprintnumbers,amsmath,amssymb]{revtex4}

\usepackage{amsmath,amssymb}
\usepackage{graphicx}
\usepackage{dcolumn}
\usepackage{bm}

\begin{document}

\title{CPT-like states in an ensemble of interacting fermions.\\
On the possibility of new mechanism of superconductivity}
\author{A. V. Taichenachev and V. I. Yudin}
\affiliation{Institute of Laser Physics SB RAS, Novosibirsk
630090, Russia, e-mail: llf@laser.nsc.ru \\ Novosibirsk State
University, Novosibirsk 630090, Russia}

\begin{abstract}
Using the standard Hamiltonian of the BCS theory, we show that in
an ensemble of interacting fermions there exists a coherent state
$|NC\rangle$, which nullifies the Hamiltonian of the
interparticle interaction. This state has an analogy with the
well-known in quantum optics coherent population trapping effect
(CPT). A possible application of such CPT-like states in the
superconductivity theory is discussed.
\end{abstract}

\pacs{42.50.Gy, 74.20.-z, 74.20.Fg, 67.57.-z}

\maketitle

\section{Introduction}

The effect of coherent population trapping (CPT) (see
\cite{Alz,Ar1,Gr,Agap,Ar2} and references therein) is one of
nonlinear interference effects. Owing to number of its
manifestations in different optical phenomena and its practical
applications CPT occupies one of leading place in modern laser
physics, nonlinear and quantum optics. For example, CPT is used
in high-resolution spectroscopy \cite{DR}, nonlinear optics of
resonance media \cite{EIT,LWI}, laser cooling \cite{VSCPT}, atom
optics and interferometry \cite{AT}, physics of quantum
information \cite{mazets96,DSPol,lightstorage}.

In the case of classical resonant field the CPT theory has been
developed for a three-state model \cite{Ar2,Gr} as well as for
multi-level systems with account for the level degeneracy
\cite{Hioe,sm,tum}. Recently we generalize this theory to the
case of an ensemble of atoms interacting with a quantized light
field \cite{Taich,Taich2}.

From the very general point of view the essence of CPT can be
formulated as follows. Consider two quantum systems (particles or
fields) $A$ and $B$. The interaction between them is described by
the Hamiltonian  $\widehat{V}_{A-B}$. Then the CPT effect occurs
when there exists a non-trivial state $|NC\rangle$, which
nullifies the interaction:
\begin{equation}\label{CPT}
\widehat{V}_{A-B}|NC\rangle=0\,.
\end{equation}
In this state, obviously, the energy exchange between the systems
$A$ and $B$ is absent. However, information correlations of the
systems can be very strong, leading to important physical
consequences. Note that if the system $A$ is equivalent to the
system $B$, then the condition (\ref{CPT}) means the absence of
the field self-interaction  or of the interparticle interaction
\begin{equation}\label{CPT1}
\widehat{V}_{A-A}|NC\rangle=0\,.
\end{equation}
From this general viewpoint the standard CPT effect in the
resonant interaction of atoms with electromagnetic field is
deciphered as follows: $A$ and $B$ is an ensemble of atoms and
resonant photons, respectively; $\widehat{V}_{A-B}=-(\hat{\bf
d}{\bf E})$ is the dipole interaction operator, and  $|NC\rangle$
is the dark state $|dark\rangle$:
\begin{equation}\label{dark}
-(\hat{\bf d}{\bf E})|dark\rangle=0\,.
\end{equation}
In the course of the interaction atoms are accumulated in the
dark state, after that they do not scatter light, and they are
not scattered by light. The information on various parameters of
the resonant field has been encoded in the state $|dark\rangle$
\cite{sm,Taich}.

Our standpoint consists in the following. The CPT principle,
expressed by (\ref{CPT}) or (\ref{CPT1}), is universal enough and
it can be manifested in various branches of physics. For the first
time such a generalized approach to CPT has been developed in our
early paper \cite{Tum}, where it is pointed out that from a
phenomenological viewpoint the CPT effect is similar to the
superconductivity. In \cite{Tum} the following comparison is
carried out: atoms and electromagnetic field from one side,
electrons and phonons form the other side. Indeed, a gas of atoms
being in the dark state $|dark\rangle$ do not interact with
photons (see eq.(\ref{dark})), similarly to electrons in a
superconducting state in solids, which are not scattered by the
phonon oscillations of a lattice. In the paper \cite{Tum} a
hypothesis on the possibility of an alternative (to the standard
BCS theory \cite{Bar}) mechanism of superconductivity has been
proposed. Namely, a quantum system of electrons and phonons
coupled by the interaction Hamiltonian $\widehat{V}_{e-phonon}$
was considered. According to \cite{Tum}, the new mechanism of
superconductivity could be based on the existence of such a state
$|NC\rangle$, which nullifies the interaction operator
$\widehat{V}_{e-phonon}$:
\begin{equation}\label{NC1}
\widehat{V}_{e-phonon}|NC\rangle=0\,,
\end{equation}
analogously to eq.(\ref{dark}). However, the explicit form of the
state  $|NC\rangle$ was not found in \cite{Tum}.

In the present paper for the standard Hamiltonian of
interparticle interaction in the BCS model \cite{Bar} we find in
explicit and analytical form a CPT-like state of the type
(\ref{CPT1}). A possible application of the obtained results in
the superconductivity (or superfluidity for $^3$He) theory is
discussed.

\section{Ensemble of fermions in a finite volume}

Consider an ensemble of fermions in a volume  $L^3$. We will use
the standard BCS Hamiltonian \cite{Bar}:
\begin{equation}\label{H}
\widehat{H}=\widehat{H}_0+\widehat{W}\,.
\end{equation}
The Hamiltonian of free particles can be written as:
\begin{equation}\label{H_0}
\widehat{H}_0=\sum_{s,{\bf k}}\varepsilon^{}_{\bf
k}\,\hat{a}^{\dag}_{s{\bf k}}\hat{a}^{}_{s{\bf k}}\,,
\end{equation}
where $\hat{a}^{\dag}_{s {\bf k}}$($\hat{a}^{}_{s {\bf k}}$) is
the creation (annihilation) operator of a particle in the state
with wavevector ${\bf k}$ and spin $s=\uparrow,\downarrow$, and
$\varepsilon^{}_{k}=(\hbar {\bf k})^2/2m$ is the energy of this
state. If the chemical potential $\mu$ is introduced into the
theory, then ($\varepsilon^{}_{\bf k}$$-$$\mu$) should be used in
(\ref{H_0}) instead of $\varepsilon^{}_{\bf k}$.

\begin{figure}[t]
\centerline{\scalebox{0.4}{\includegraphics{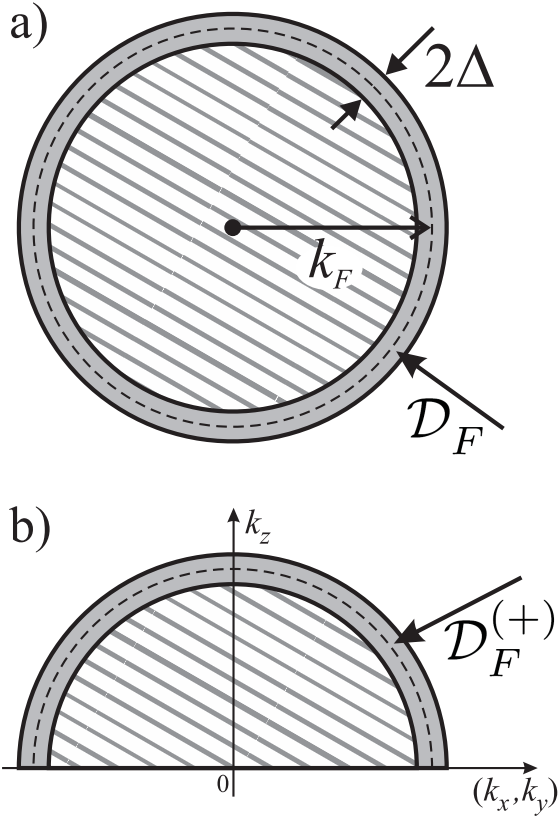}}}\caption{Illustrations:
a) thin spherical layer ${\cal D}_F$ with the width 2$\Delta$
around the Fermi surface with the radius $k_F$; b) upper
hemispherical layer ${\cal D}_F^{(+)}$ ñ $k_z$$>$0.}
\end{figure}

The interaction between particles is described by the Hamiltonian
coupling particles with opposite momenta and spin:
\begin{eqnarray}\label{W}
&&\widehat{W}=\frac{g}{L^3}\sum_{{\bf k}_{1,2}\in {\cal D}_F}G({\bf k}_1,{\bf k}_2)\,\hat{a}^{\dag}_{\uparrow {\bf
k}_1}\hat{a}^{\dag}_{\downarrow -{\bf k}_1} \hat{a}^{}_{\downarrow -{\bf k}_2}\hat{a}^{}_{\uparrow {\bf k}_2}\,,\\
&&{\cal D}_F :\;\; k_F-\Delta\le |{\bf k}|\le
k_F+\Delta\,.\nonumber
\end{eqnarray}
Only particles with wavevectors in the thin layer of the width
$2\Delta$ around the Fermi surface (see in Fig.1a), having the
radius $k_F$ ($\Delta\ll k_F$), are involved in the interaction.
This subset in the wavevector space will be referred to as ${\cal
D}_F$. If even one of the vectors ${\bf k}_{1,2}$ does not belong
to the subset ${\cal D}_F$, then $G$(${\bf k}_1,$${\bf k}_2$)=0.
The sign of the interaction constant $g$ in (\ref{W}) governs the
attraction ($g$$<$0) or repulsion ($g$$>$0) between particles.
The formfactor $G$(${\bf k}_1,$${\bf k}_2$) obeys to the general
symmetry condition
\begin{equation}\label{G}
G({\bf k}_1,{\bf k}_2)=G(-{\bf k}_1,{\bf k}_2)=G({\bf k}_1,-{\bf k}_2)\,.
\end{equation}
It is usually assumed that $G$(${\bf k}_1,$${\bf k}_2$)=1 at
${\bf k}_{1,2}$$\in$${\cal D}_F$. Recall that, according to the
standard conception, the model Hamiltonian (\ref{W}) is
determined by the interaction of electrons with the phonons of
lattice and Coulomb repulsion between electrons.

It turns out that the operator (\ref{W}) allows the existence of
the CPT-like state $|NC\rangle$, obeying the condition
\begin{equation}\label{W_NC1}
\widehat{W}|NC\rangle=0\,.
\end{equation}
Let us build up this state. Consider first the following operator
construction:
\begin{equation}\label{constr}
\hat{b}^{\dag}_{\bf k}(\lambda)=\hat{a}^{\dag}_{\uparrow {\bf k}}\hat{a}^{\dag}_{\downarrow -{\bf k}}+\lambda\,
\hat{a}^{\dag}_{\uparrow -{\bf k}}\hat{a}^{\dag}_{\downarrow {\bf k}}\,,
\end{equation}
which generates two-particle coupled states with opposite
wavevectors ${\bf k}$ and $-{\bf k}$, and with zero total spin;
the parameter $\lambda$ is arbiarary number. Using the standard
anticommutator rules for fermionic operators $\hat{a}^{\dag}_{s
{\bf k}}$ and $\hat{a}^{}_{s {\bf k}}$, and the property
(\ref{G}), we calculate the commutator:
\begin{eqnarray}\label{comm1}
&&\left[\widehat{W},\,\hat{b}^{\dag}_{\bf k}(\lambda)\right]=\frac{g}{L^3}\sum_{{\bf k}_{1}\in {\cal D}_F}G({\bf
k}_1,{\bf k})\,\hat{a}^{\dag}_{\uparrow
{\bf k}_1}\hat{a}^{\dag}_{\downarrow -{\bf k}_1}\\
&&(1+\lambda-\lambda\hat{a}^{\dag}_{\uparrow -{\bf k}}\hat{a}^{}_{\uparrow -{\bf k}}-
\lambda\hat{a}^{\dag}_{\downarrow {\bf k}}\hat{a}^{}_{\downarrow {\bf k}}-\hat{a}^{\dag}_{\uparrow
{\bf k}}\hat{a}^{}_{\uparrow {\bf k}}- \hat{a}^{\dag}_{\downarrow -{\bf k}}\hat{a}^{}_{\downarrow
-{\bf k}})\nonumber .
\end{eqnarray}
As is seen, when $\lambda=-1$ this commutator has the specific
form, where all summands are finished by the annihilation
operators $\hat{a}^{}_{\uparrow \pm{\bf k}}$ and
$\hat{a}^{}_{\downarrow \pm{\bf k}}$. Therefore we define now the
basic operator construction $\hat{\gamma}^{\dag}_{\bf k}$:
\begin{equation}\label{base}
\hat{\gamma}^{\dag}_{\bf k}\equiv\hat{b}^{\dag}_{\bf k}(-1)=\hat{a}^{\dag}_{\uparrow {\bf
k}}\hat{a}^{\dag}_{\downarrow -{\bf k}}-\hat{a}^{\dag}_{\uparrow -{\bf k}}\hat{a}^{\dag}_{\downarrow {\bf k}}\,,
\end{equation}
for which the commutator (\ref{comm1}) takes the form:
\begin{eqnarray}\label{comm2}
&&\left[\widehat{W},\,\hat{\gamma}^{\dag}_{\bf k}\right]=\frac{g}{L^3}\sum_{{\bf k}_{1}\in {\cal D}_F}G({\bf
k}_1,{\bf k})\,\hat{a}^{\dag}_{\uparrow
{\bf k}_1}\hat{a}^{\dag}_{\downarrow -{\bf k}_1}\nonumber\\
&&(\hat{a}^{\dag}_{\uparrow -{\bf k}}\hat{a}^{}_{\uparrow -{\bf
k}}+ \hat{a}^{\dag}_{\downarrow {\bf k}}\hat{a}^{}_{\downarrow
{\bf k}}-\hat{a}^{\dag}_{\uparrow {\bf k}}\hat{a}^{}_{\uparrow
{\bf k}}- \hat{a}^{\dag}_{\downarrow -{\bf
k}}\hat{a}^{}_{\downarrow -{\bf k}})\,.\quad
\end{eqnarray}
This expression is crucial for the building up the CPT-like state
(\ref{W_NC1}).

It is worth to note that due to the obvious relationship
\begin{equation}\label{k-k}
\hat{\gamma}^{\dag}_{-\bf k}=-\hat{\gamma}^{\dag}_{\bf k}
\end{equation}
the operators $\hat{\gamma}^{\dag}_{\bf k}$, defined on the
spherical layer ${\bf k}$$\in$${\cal D}_F$, are not independent.
Therefore instead of the subset ${\cal D}_F$ we define the upper
hemispherical layer ${\cal D}_F^{(+)}$ (see Fig.1b), consisting
of vectors ${\bf k}$$\in$${\cal D}_F$ with positive projections
on the axis $Oz$ ($k_z>0$) only. Now the operators
$\hat{\gamma}^{\dag}_{\bf k}$, defined for vectors ${\bf
k}$$\in$${\cal D}_F^{(+)}$ will be independent.

Consider the operator construction
\begin{equation}\label{Psi}
\widehat{\Psi}_{NC}=\prod_{{\bf k}\in {\cal D}_F^{(+)}}\hat{\gamma}^{\dag}_{\bf k}\,,
\end{equation}
which acts on the upper subset  ${\cal D}_F^{(+)}$ (for each
${\bf k}$ the operator $\hat{\gamma}^{\dag}_{\bf k}$ is used, at
most, once). Obviously, the order of multipliers can be arbitrary,
because [$\hat{\gamma}^{\dag}_{\bf
k}$$,\,$$\hat{\gamma}^{\dag}_{\bf k'}$]=0. Let us factor out
arbitrary operator  $\hat{\gamma}^{\dag}_{\bf k'}$ in (\ref{Psi})
from the product $\Pi$ and then act by the operator $\widehat{W}$
on $\widehat{\Psi}_{NC}$:
\begin{equation}\label{W_Psi}
\widehat{W}\widehat{\Psi}_{NC}=\widehat{W}\hat{\gamma}^{\dag}_{\bf k'}\prod_{{\bf k}\neq {\bf k'}
}\hat{\gamma}^{\dag}_{\bf k}=\left(\hat{\gamma}^{\dag}_{\bf k'}\widehat{W}+[\widehat{W},\,\hat{\gamma}^{\dag}_{\bf
k'}]\right)\prod_{{\bf k}\neq {\bf k'} }\hat{\gamma}^{\dag}_{\bf k}.
\end{equation}
Since under the sign $\Pi$ in (\ref{W_Psi}) the creation
operators with wavevectors $\pm{\bf k'}$ are absent, then, as is
follows from eq.(\ref{comm2}), the commutator
$[\widehat{W},\,\hat{\gamma}^{\dag}_{\bf k'}]$ can be moved to
the right side through the product $\Pi$. As a result, the
expression (\ref{W_Psi}) can be written as:
\begin{equation}\label{W_Psi1}
\widehat{W}\widehat{\Psi}_{NC}=\hat{\gamma}^{\dag}_{\bf k'}\widehat{W}\prod_{{\bf k}\neq {\bf k'}
}\hat{\gamma}^{\dag}_{\bf k}+\left(\prod_{{\bf k}\neq {\bf k'} }\hat{\gamma}^{\dag}_{\bf
k}\right)[\widehat{W},\,\hat{\gamma}^{\dag}_{\bf k'}]\,.
\end{equation}
Let us consider also the operator construction
\begin{equation}\label{Ferm1}
\widehat{\Phi}(\Delta)=\prod_{|{\bf k}|< (k_F-\Delta)}\hat{a}^{\dag}_{\uparrow{\bf
k}}\hat{a}^{\dag}_{\downarrow{\bf k}}\,,
\end{equation}
which, acting on the vacuum $|0\rangle$, generates the state,
corresponding to the completely occupied sphere with the radius
($k_F-\Delta$) in the wavevector space (in Fig.1a it corresponds
to the inner sphere shaded by skew lines). The following
commutator relationships are evident:
\begin{equation}\label{comm_F}
[\widehat{W},\, \widehat{\Phi}(\Delta)]=0;\quad \left[ [\widehat{W},\,\hat{\gamma}^{\dag}_{\bf k'}],\,
\widehat{\Phi}(\Delta)\right]=0\quad ({\bf k'}\in {\cal D}_F)\,,
\end{equation}
because in the operator  $\widehat{\Phi}(\Delta)$ (see
(\ref{Ferm1})) only the wavevectors  $|{\bf
k}|$$<$$(k_F$$-\Delta)$ are used. These vectors do not belong to
the upper layer ${\cal D}_F$ where the operators $\widehat{W}$ and
$\hat{\gamma}^{\dag}_{\bf k'}$ act.

Let us prove that the state $|NC\rangle$, nullifying the
interaction (\ref{W_NC1}), has the form
\begin{equation}\label{NC_constr}
|NC\rangle=\widehat{\Psi}_{NC}\widehat{\Phi}(\Delta)|0\rangle\,.
\end{equation}
Acting on this state by the operator, and taking into account the
relationships (\ref{W_Psi1}) and (\ref{comm_F}), one can obtain:
\begin{eqnarray}\label{W_NCconstr}
&&\widehat{W}\widehat{\Psi}_{NC}\widehat{\Phi}(\Delta)|0\rangle =\hat{\gamma}^{\dag}_{\bf
k'}\widehat{W}\left(\prod_{{\bf k}\neq {\bf k'}
}\hat{\gamma}^{\dag}_{\bf k}\right)\widehat{\Phi}(\Delta)|0\rangle+\nonumber\\
&&\left(\prod_{{\bf k}\neq {\bf k'} }\hat{\gamma}^{\dag}_{\bf
k}\right)\widehat{\Phi}(\Delta)\,[\widehat{W},\,\hat{\gamma}^{\dag}_{\bf k'}]\,|0\rangle\,.
\end{eqnarray}
However, since the commutator (\ref{comm2}) is finished from the
right side by the annihilation operators, then
[$\widehat{W}$$,\,$$\hat{\gamma}^{\dag}_{\bf k'}$]$|0\rangle=0$.
Thus, from (\ref{W_NCconstr}) we have
\begin{equation}\label{W_NCcomm}
\widehat{W}\widehat{\Psi}_{NC}\widehat{\Phi}(\Delta)|0\rangle =\hat{\gamma}^{\dag}_{\bf
k'}\widehat{W}\left(\prod_{{\bf k}\neq {\bf k'} }\hat{\gamma}^{\dag}_{\bf
k}\right)\widehat{\Phi}(\Delta)|0\rangle\,,
\end{equation}
From this equation we see that it is possible to change the
sequence order of $\widehat{W}$ and any operator
$\hat{\gamma}^{\dag}_{\bf k}$. Proceeding this consideration step
by step and taking into account (\ref{comm_F}), we obtain
eventually:
\begin{eqnarray}\label{W_NC_fin}
&&\widehat{W}|NC\rangle\equiv
\widehat{W}\widehat{\Psi}_{NC}\widehat{\Phi}(\Delta)|0\rangle
=\widehat{\Psi}_{NC}
\widehat{W}\widehat{\Phi}(\Delta)|0\rangle=\nonumber\\
&&\widehat{\Psi}_{NC}\widehat{\Phi}(\Delta)\widehat{W}|0\rangle=0\,.
\end{eqnarray}
Here the last transformation to zero is obvious, because the
operator $\widehat{W}$ (see (\ref{W})) is finished from the right
side by the annihilation operators $\hat{a}^{}_{s\pm {\bf k}}$.
Thus, we prove rigorously that the state (\ref{NC_constr})
nullifies the interparticle interaction (scattering), i.e. it
obeys the equation (\ref{W_NC1}).

It should be noted the presence of the construction
$\widehat{\Phi}(\Delta)$ in (\ref{NC_constr}) is necessary from
the physical point of view, since the form of the interaction
Hamiltonian (\ref{W}), according to \cite{Bar}, is a consequence
of almost completely  occupied Fermi sphere. Thus, physically
significant states should differ from the ideal Fermi state
$|F\rangle$:
\begin{equation}\label{Fermi}
|F\rangle=\left(\prod_{|{\bf k}|\le k_F}\hat{a}^{\dag}_{\uparrow{\bf k}}\hat{a}^{\dag}_{\downarrow{\bf
k}}\right)|0\rangle
\end{equation}
only in a small region nearby the Fermi sphere. For the state
(\ref{NC_constr}) this difference is described by the construction
$\widehat{\Psi}_{NC}$ (\ref{Psi}), acting in the thin layer ${\cal
D}_F$ around the Fermi surface in the wavevector space.

The state $|NC\rangle$ is an eigenstate for the unperturbed
Hamiltonian $\widehat{H}_0$ and, consequently, for the total
Hamiltonian $\widehat{H}$:
\begin{eqnarray}\label{H_NC}
&&\widehat{H}\,|NC\rangle=\widehat{H}_0\,|NC\rangle=E_{NC}\,|NC\rangle\,,\nonumber\\
&&E_{NC}=\frac{3}{5}\,{\cal E}_F{\cal N}\left\{
1+10\left(\frac{\Delta}{k_F}\right)^2+5\left(\frac{\Delta}{k_F}\right)^4\right\},\qquad
\end{eqnarray}
where ${\cal E}_F$=$(\hbar k_F)^2$/2$m$ is the Fermi energy and à
${\cal N}$ is the number of particles in ensemble. In the theory
with chemical potential $\mu$ we should add $-\mu$${\cal N}$ to
the value $E_{NC}$ in (\ref{H_NC}).

As to the construction $\widehat{\Psi}_{NC}$, the occupation of
all the thin layer ${\cal D}_F$ in (\ref{Psi}) is dictated by the
conservation of particle number. Indeed, as it follows from
(\ref{base}), each operator $\hat{\gamma}^{\dag}_{\bf k}$
describes the distribution of two electrons among the four states
$|$$\uparrow$,${\bf k}\rangle$, $|$$\downarrow$,${\bf k}\rangle$,
$|$$\uparrow$,$-{\bf k}\rangle$, $|$$\downarrow$,$-{\bf
k}\rangle$. Because of this, in order to distribute all
electrons, which at the density packing (into Fermi sphere) were
located in the layer ($k_F$$-\Delta$)$\leq$$k$$\leq$$k_F$, we
need in a doubled volume in the wavevector space. In the case
$\Delta$$\ll$$k_F$ practically the whole thin layer ${\cal D}_F$
(see in Fig.1a) corresponds to a such double volume, fro which
($k_F$$-\Delta$)$\leq$$k$$\leq$($k_F$$+\Delta$). If the particle
number conservation is not taken into account, then an arbitrary
number of different operators $\hat{\gamma}^{\dag}_{\bf k}$ can
be used in the construction (\ref{Psi}).

Note that the ground state in the BCS theory \cite{Bar} can be
written in the form
\begin{equation}\label{BCS}
|BCS\rangle=\widehat{\Psi}_{BCS}\widehat{\Phi}(\Delta)|0\rangle\,
\end{equation}
with the operator construction
\begin{equation}\label{psi_BCS}
\widehat{\Psi}_{BCS}=\prod_{{\bf k}\in {\cal D}_F}\left\{ u({\bf k})+v({\bf k})\hat{a}^{\dag}_{\uparrow{\bf
k}}\hat{a}^{\dag}_{\downarrow -{\bf k}}\right\}\,,
\end{equation}
where $u({\bf k})$ and $v({\bf k})$ are variational coefficients
(they are coefficients in the Bogolyubov transformation).

Let us discuss some basic properties of the state $|NC\rangle$,
which are quite different from those of the state $|BCS\rangle$
in the BCS theory:\\
I. $|NC\rangle$ is an eigenstate for the particle number operator
$\widehat{N}$=$\sum_{s,{\bf k}}\hat{a}^{\dag}_{s{\bf
k}}\hat{a}^{}_{s{\bf k}}$:
\begin{equation}\label{N}
\widehat{N}|NC\rangle={\cal N}|NC\rangle\,.
\end{equation}
II. The state $|NC\rangle$ is an eigenstate for the total
momentum operator $\widehat{\bf P}$=$\sum_{{\bf k},s}(\hbar{\bf
k})\hat{a}^{\dag}_{s{\bf k}}\hat{a}^{}_{s{\bf k}}$. For example,
if the Fermi sphere is constructed around the wavevector ${\bf
K}$, then we have:
\begin{equation}\label{P_NC}
\widehat{\bf P}|NC\rangle={\cal N}(\hbar {\bf K})|NC\rangle\,,
\end{equation}
i.e. this state corresponds to a free flow of particles. (The
consideration above dealt with the particular case ${\bf K}$=0,
but the generalization to arbitrary ${\bf K}$ is almost
elementary.)\\
III. $|NC\rangle$ does not depend on the value and sign of the
coupling constant $g$, i.e. it exists in both cases of weak and
strong coupling, and for the case of interparticle repulsion.\\
IV. In the BCS theory the product $\hat{a}^{\dag}_{\uparrow{\bf
k}}\hat{a}^{\dag}_{\downarrow -{\bf k}}$ corresponds to the
operator of creation of scalar quasiparticle (Cooper pair), while
in our case the construction $\hat{\gamma}^{\dag}_{\bf k}$ is
related to a coupled pair. This construction, as it follows from
(\ref{base}), generates an entangled state with respect to the
spin and translational degrees of freedom.

Note, that a role of entanglement in strongly correlated spin
systems (in particular for fermions in BCS) was investigated in
\cite{Martin1,Martin2}.

\section{Conclusion}
We have found the CPT-like state $|NC\rangle$ in an ensemble of
fermions with spin 1/2. This state has an analogy with the CPT
effect and it nullifies the interparticle interaction operator of
the standard BCS theory.

Evidently, the states $|NC\rangle$ constitutes a special class of
eigenstates of the total Hamiltonian $\widehat{H}$ due to the
independency on the coupling constant $g$, while, undoubtedly,
there exist other eigenstates with a nontrivial analytical
$g$-dependence of the energy $E$($g$). Because of this a question
about the physical realization of the state $|NC\rangle$ requires
a separate consideration. In the case of interparticle attraction
($g<$0) the energy $E_{NC}$ for the CPT-like state lies above the
ground-state energy of the BCS theory. However, in the theory
with interparticle repulsion ($g>$0) it is possible, in principle,
that the state $|NC\rangle$ will be the ground state, because
other states acquire a positive increment to the energy. Note
that due to the Coulomb repulsion (or excitonic mechanism of
interaction) between electrons the theory with interparticle
repulsion has equal chances for the realization relative to the
theory with attraction.

It should be stressed that, according to the property II, the
state $|NC\rangle$ corresponds to a free (non-dissipative)
particle flow (\ref{P_NC}) and it is a superconduting (or
superfluid for $^3$He) state just due to its nature (even without
references to energy gap). From this point of view the presence
of energy gap is necessary first of all for the stability of this
state.

Although the problem on the realization of CPT-like states
remains open, nevertheless the presented study argues for a
principal possibility of alternative approaches, even in the
framework of standard BCS Hamiltonian (\ref{W}). It looks
interesting, from our viewpoint, due to difficulties with
construction of a complete theory of the high-temperature
superconductivity. This, in its turn, stimulates the search of
various alternatives for the BCS theory (see, for example, review
\cite{Ginz}). The approach outlined in the present paper is
another alternative based on an analogy with the CPT effect. It
is advisable to study this alternative in detail in the future.
Besides, it is possible that the CPT-like states will find their
applications regardless to the superconductivity theory.

This work was supported by RFBR (grants \# 05-02-17086,
04-02-16488, 05-08-01389, 07-02-01230).

\end{document}